\documentclass[lettersize,journal]{IEEEtran}
\IEEEoverridecommandlockouts

\usepackage{cite}
\usepackage[numbers,sort&compress]{natbib}
\usepackage{amsmath,amssymb,amsfonts}
\usepackage{amsmath, amssymb, amsthm, mathrsfs}

\usepackage{algorithmic}
\usepackage{algorithm}
\usepackage{graphicx}
\usepackage{url}
\usepackage[caption=false,font=normalsize,labelfont=sf,textfont=sf]{subfig}
\usepackage{textcomp}
\usepackage{stfloats}
\usepackage{verbatim}
\usepackage{caption}
\usepackage{subcaption}
\usepackage{textcomp}
\usepackage{xcolor}
\usepackage{hyperref}
\hypersetup{
    colorlinks=true,
    linkcolor=blue,
    citecolor=blue,
    urlcolor=blue
}

\def\BibTeX{{\rm B\kern-.05em{\sc i\kern-.025em b}\kern-.08em
    T\kern-.1667em\lower.7ex\hbox{E}\kern-.125emX}}
        
\begin{document}

\UseRawInputEncoding

\title{Multi-Satellite Beam Hopping and Power Allocation Using Deep Reinforcement Learning\\
}

\author{{Xia Xie,
Kexin Fan,
Wenfeng Deng,
Nikolaos Pappas,~\IEEEmembership{Senior Member,~IEEE},
Qinyu Zhang,~\IEEEmembership{Senior Member,~IEEE}

}

\thanks{This work was supported in part by the National Natural Science Foundation of China (NSFC) under Grant 62027802, and in part by the Shenzhen Science and Technology Program under Grant KQTD20210811090116029. An earlier version of this work was presented in part at the IEEE VTC2024-Fall \cite{r0}. (Corresponding author: Qinyu Zhang.) }
\thanks{X. Xie, K. Fan, and Q. Zhang are with Guangdong Provincial Key Laboratory of Aerospace Communication and Networking Technology, Harbin Institute of Technology (Shenzhen), Shenzhen 518055, China (e-mail: xiexia@stu.hit.edu.cn; kexin.fan1@foxmail.com; zqy@hit.edu.cn).}

\thanks{W. Deng is with the Northwest Institute of Nuclear Technology, Xi'an 710000, China (e-mail:785329245@qq.com). }

\thanks{Nikolaos Pappas is with the Department of Computer and Information Science at  Link{\"o}ping University, Sweden, (e-mail: nikolaos.pappas@liu.se). }

}

\maketitle
\begin{abstract}

In non-geostationary orbit (NGSO) satellite communication systems, effectively utilizing beam hopping (BH) technology is crucial for addressing uneven traffic demands. However, optimizing beam scheduling and resource allocation in multi-NGSO BH scenarios remains a significant challenge. This paper proposes a multi-NGSO BH algorithm based on deep reinforcement learning (DRL) to optimize beam illumination patterns and power allocation. By leveraging three degrees of freedom (i.e., time, space, and power), the algorithm aims to optimize the long-term throughput and the long-term cumulative average delay (LTCAD). The solution is based on proximal policy optimization (PPO) with a hybrid action space combining discrete and continuous actions. Using two policy networks with a shared base layer, the proposed algorithm jointly optimizes beam scheduling and power allocation. One network selects beam illumination patterns in the discrete action space, while the other manages power allocation in the continuous space. Simulation results show that the proposed algorithm significantly reduces LTCAD while maintaining high throughput in time-varying traffic scenarios. Compared to the four benchmark methods, it improves network throughput by up to $8.9\%$ and reduces LTCAD by up to $69.2\%$.

\end{abstract}


\section{Introduction}
\subsection{Background and Motivation}
\IEEEPARstart{T}{he} goal of future communication networks is to achieve global coverage and provide high-speed data transmission services \cite{r1}. The space-ground integrated network has become an inevitable trend in achieving this goal. With the rapid development of low earth orbit (LEO) constellations such as OneWeb, Starlink, and Telesat, the connection between satellite networks and ground networks has been further strengthened \cite{r2},\cite{r3}. However, the payload and resources of a single satellite are limited and cannot fully meet the large number of user demands on the ground \cite{r4}. Therefore, non-geostationary orbit (NGSO) satellites should collaborate to achieve seamless connectivity across multiple coverage areas \cite{r5},\cite{r6}. Compared with the wide-area coverage of satellites, satellites using multi-beam technology can effectively address the challenge of uneven distribution of terrestrial traffic \cite{r7}. The quality of user communications is significantly improved by generating multiple high-gain spot beams and focusing the energy precisely on the target area. In traditional fixed-beam multi-beam satellites, the beam directions are predetermined during the design phase and remain constant throughout the operation. While this design ensures coverage of a specific area, it does not fully leverage the spatial flexibility of the beam. With the advancement of phased array antenna technology, beam hopping (BH) technology has gradually been introduced into multi-beam satellite systems. This technology enables beams to switch between different ground areas quickly, thereby improving the spatial freedom of the beam. Currently, BH technology is employed in Starlink and DVB-S2X standards, and its widespread adoption in NGSO constellations is anticipated to be a key trend in the future \cite{r8}. However, in a multi-satellite multi-beam system, the inevitable interference between neighboring beams and the dynamic link changes due to the fast movement of satellites can increase the complexity of resource allocation and beam management. Therefore, designing effective beam illumination patterns and resource allocation strategies to adapt to uneven traffic demands is crucial for BH communication systems.

\subsection{Related Work}
Recently, there has been a growing focus on beam illumination patterns and resource allocation. Numerous algorithms have been developed specifically to optimize beam illumination patterns. In \cite{r9}, an iterative algorithm was introduced to minimize Co-Channel Interference (CCI) and maximize Signal-to-Interference-plus-Noise Ratio (SINR), aiming to maximize throughput by assigning beams to links with the highest SINR. Similarly, methods based on Viterbi and greedy algorithms were proposed in \cite{r10} to optimize BH patterns by minimizing CCI between adjacent beams. The authors in \cite{r11} proposed a dynamic beam illumination schedule with selective precoding, using an interference-based penalty function to address high traffic demands. A combined learning and optimization approach was proposed in \cite{r12} to identify BH patterns that maximize user satisfaction. These studies mainly focus on BH illumination patterns. The authors in \cite{r14} proposed a BH scheduling and bandwidth allocation method that considers interference threshold distance and beam service priority for resource allocation. In \cite{r15}, a joint scheme for all-digital beamforming and illumination patterns was developed using random search and fractional programming. The authors in \cite{r16} proposed a multi-domain decoupled allocation algorithm that utilizes relaxation and convex approximation techniques in an inner-outer loop framework. This approach optimizes resource allocation in LEO satellite systems by minimizing energy consumption and unmet capacity. While these algorithms effectively improve resource utilization, they primarily focus on single-satellite scenarios. For multi-satellite scenarios, the authors in \cite{r17} proposed a method using a greedy algorithm and convex optimization for the beam scheduling and power allocation in LEO and geostationary orbit (GEO) coexistence satellites. In \cite{r18}, the authors employed an adaptive particle algorithm to tackle the BH problem within a dual-satellite cooperative transmission network. The authors in \cite{r19}  proposed a multi-satellite BH algorithm based on load balancing and interference avoidance, decomposing the problem into load balancing, single-satellite BH pattern design, and multi-satellite interference avoidance. These studies mainly focused on BH and resource allocation but were based on pre-planned beam scheduling, assuming a prior knowledge of traffic demands in each cell. To meet the dynamically changing traffic demands, traffic-driven BH scheduling methods have attracted much attention, which are adjusting the beam illumination according to real-time traffic and user needs. In \cite{r21}, a weighted greedy strategy was proposed for determining beam illumination patterns and an improved genetic algorithm to optimize power and bandwidth allocation. The authors in \cite{r23} tackled the challenge of large solution spaces in resource scheduling models using an enhanced artificial bee colony algorithm. The authors in \cite{r24} proposed a resource allocation algorithm to maximize user service weight gain. This algorithm prioritizes resources for higher-weight cells by considering traffic demand and delay sensitivity.
In \cite{r25}, a heuristic beam scheduling algorithm and a maximum weighted algorithm were proposed to match traffic distribution and latency requirements better. The authors in \cite{r26} transformed the dynamic beam placement problem into a p-center problem, aiming to minimize the number of beam positions needed to cover all users. These algorithms are driven by traffic demand and make decisions based on real-time traffic dynamics. However, traditional methods are often prone to local optimal solutions when dealing with complex problems.

Deep reinforcement learning (DRL) is an effective decision-making tool widely regarded as a powerful method for solving such problems due to its fast convergence and adaptability in complex decision-making problems. The authors in \cite{r27} proposed a method based on Hull moving average and Q-learning to allocate beam bandwidth. In \cite{r28}, The authors proposed a DRL-based algorithm to optimize the BH scheduling and coverage control jointly. In \cite{r29}, the authors proposed a dual-loop learning method for optimal BH in DVB-S2X satellites to maximize throughput and minimize latency. In \cite{r30}, a parameterized RL-based joint optimization method was introduced to adjust beam patterns and power allocation. In \cite{r31}, the authors propose a joint optimization approach based on parameterized reinforcement learning to simultaneously regulate BH and power allocation.
The authors in \cite{r32} utilized a genetic algorithm combined with DRL to optimize decisions in dynamic satellite environments. In \cite{r33}, a cooperative multi-agent DRL framework was developed, where each agent manages either beam illumination or bandwidth allocation. The authors in \cite{r34} introduced a joint BH scheduling and power optimization algorithm, treating beam scheduling as a potential game to achieve Nash equilibrium while optimizing power allocation with a penalty function method. However, these optimization algorithms are primarily designed for single-satellite scenarios, making it challenging to ensure a globally optimal solution when applied directly to multi-satellite systems. Furthermore, in multi-satellite scenarios, the complexity of collaboration and interference management significantly adds to the difficulty of algorithm design.

Overall, three main issues remain in the above existing works, which are elaborated as follows.
\begin{itemize}
\item[$\bullet$] Most investigations on BH primarily focused on single satellite scenarios, particularly GEO satellites \cite{r9},\cite{r10},\cite{r11},\cite{r12},\cite{r14},\cite{r15},\cite{r16},\cite{r23},\cite{r24},\cite{r25},\cite{r26},
\cite{r27},\cite{r28},\cite{r29},\cite{r30},\cite{r31},\cite{r32},\cite{r33},\cite{r34}. Compared to GEO satellite scheduling, solutions for multi-NGSO satellite scenarios face more significant challenges, such as interference management, inter-satellite coordination, and the rapid movement of satellites. These challenges are further complicated by the limited payload capacity of each satellite and the added complexity of managing numerous satellites.
\item[$\bullet$] Most existing BH research assumed that the traffic demand for each cell is known in advance \cite{r9},\cite{r10},\cite{r11},\cite{r12},\cite{r14},\cite{r15},\cite{r16},\cite{r17},\cite{r18},\cite{r19}. While these methods are effective for managing stable and predictable traffic demands, they lack the flexibility needed to respond to sudden and dynamic traffic changes. This limitation can lead to inefficient resource utilization and prolonged queuing delays \cite{r20}.
\item[$\bullet$] 
Most existing works were based on traditional methods 
\cite{r9},\cite{r10},\cite{r11},\cite{r12},\cite{r14},\cite{r15},\cite{r16},\cite{r17},\cite{r18},\cite{r19},\cite{r21},\cite{r23},
\cite{r24},\cite{r25},\cite{r26}. For complex optimization problems, traditional algorithms often attempt to decouple and address joint optimization issues separately. However, this approach makes it challenging to achieve truly optimal solutions, particularly in scenarios that require an integrated consideration of multiple factors.
\end{itemize}

\subsection{Main Contributions}

In this paper, we propose a multi-NGSO BH algorithm based on DRL to optimize beam illumination patterns and power allocation jointly. We first construct a multi-satellite BH model that comprehensively considers resource allocation among satellites as well as various system performance indicators. To achieve effective joint decision-making, we employ an algorithm based on proximal policy optimization (PPO), which employs a hybrid action space, including discrete and continuous action spaces. Specifically, the algorithm uses two participant networks with a shared base layer, one of which is responsible for determining the beam illumination pattern and the other for regulating the power allocation. The DRL method proposed in this paper fully utilizes the three degrees of freedom of the beam: time, space, and power. The algorithm can flexibly respond to time-varying and uneven traffic demands by dynamically adjusting these degrees of freedom. Finally, we conduct simulations in different scenarios to evaluate the long-term network throughput and delay performance of the proposed algorithm. Compared with the other four benchmark methods, our algorithm demonstrates significant performance gains in network throughput efficiency and queuing delay.
The main contributions of this paper are summarized below:
\begin{itemize}
\item[$\bullet$] This paper investigates the BH and power allocation problem in multi-NGSO scenarios. To maximize system throughput and minimize the system queuing delays, we formulate the problem as a multi-objective optimization problem.
\item[$\bullet$] 
We formulate the problem as a Markov decision process (MDP) and use the DRL algorithm. For joint optimization, a hybrid action space combines discrete and continuous action spaces. The proposed algorithm makes full use of the degrees of freedom of time, space, and power. 

\item[$\bullet$] 
We evaluate long-term network throughput and latency performance by conducting simulations under different scenarios. We verify the superiority of our algorithm by comparing it with four benchmark methods.
\end{itemize}

The remainder of this paper is organized as follows. Section II describes the system model in a multi-NGSO scenario. In Section III, the problem formulation is presented. Section IV introduces the traffic-driven beam hopping (BH) scheduling and power allocation method based on DRL. Section V presents the simulation results and performance analysis. Finally, Section VI concludes the paper with a summary of key findings.
For convenience, the detailed notations and definitions used in this paper are summarized in Table \ref{tab1}.

\begin{table}[!http]
	\vspace{-1.0em}
	\caption{Notations and Definitions\label{tab1}}
	\begin{center}
		\begin{tabular*}{\linewidth}{@{}ll@{}}			
			\hline
			Notations & Definitions\\
			\hline
			$N_s$ & Number of satellites\\
			$N_c$ &  Number of cells\\
			$K$  & Number of beams\\
			$V_i$& The set of cells covered by the $i$-th satellite\\
			$T_{slot}$ & The duration of each time slot\\
			$\rho_t^n$ & The traffic arrival rate of cell $n$ in time slot $t$\\
			$Q_t^n$ & The total traffic packets stored in the queue $n$ in time slot $t$\\
			$\phi_{t,l}^n$ & Data packets that have waited for $l$ time slots at the queue $n$\\
			$x_{i,n}^t$ & whether virtual cell $n$ is illuminated
			by satellite $i$ at time slot $t$\\
			$P_{tot}$ & The total satellite power\\
			$B$ & The Total satellite bandwidth\\
			$o$ & The packet size\\
			$T_{rx}$ & The noise temperature of the receiver\\
			$h_{i,n}^t$ & Channel coefficient between satellite $i$ and cell $n$ at time slot $t$\\
			$\theta_{i,n}^t$ & Off-axis angle between satellite $i$ to cell $n$ at time slot $t$\\
			$G_t(\theta_{i,n}^t)$ & Transmit antenna gain from satellite $i$ to cell $n$ at time slot $t$\\
			$G_r(\theta_{i,n}^t)$ & Receive antenna gain of cell $n$ from satellite $i$ at time slot $t$\\
			$l_{i,n}^t$ & The distance between satellite $i$ and the cell $n$ at time slot $t$\\
			$L_{i,n}^t$  & The path loss from satellite $i$ to cell $n$ at time slot $t$\\
			\hline
		\end{tabular*}
	\end{center}
	\vspace{-0.5em}
\end{table}

\section{System Model}

\begin{figure}[!htbp]
\vspace{0em}
\centerline{\includegraphics[width=0.5\textwidth]{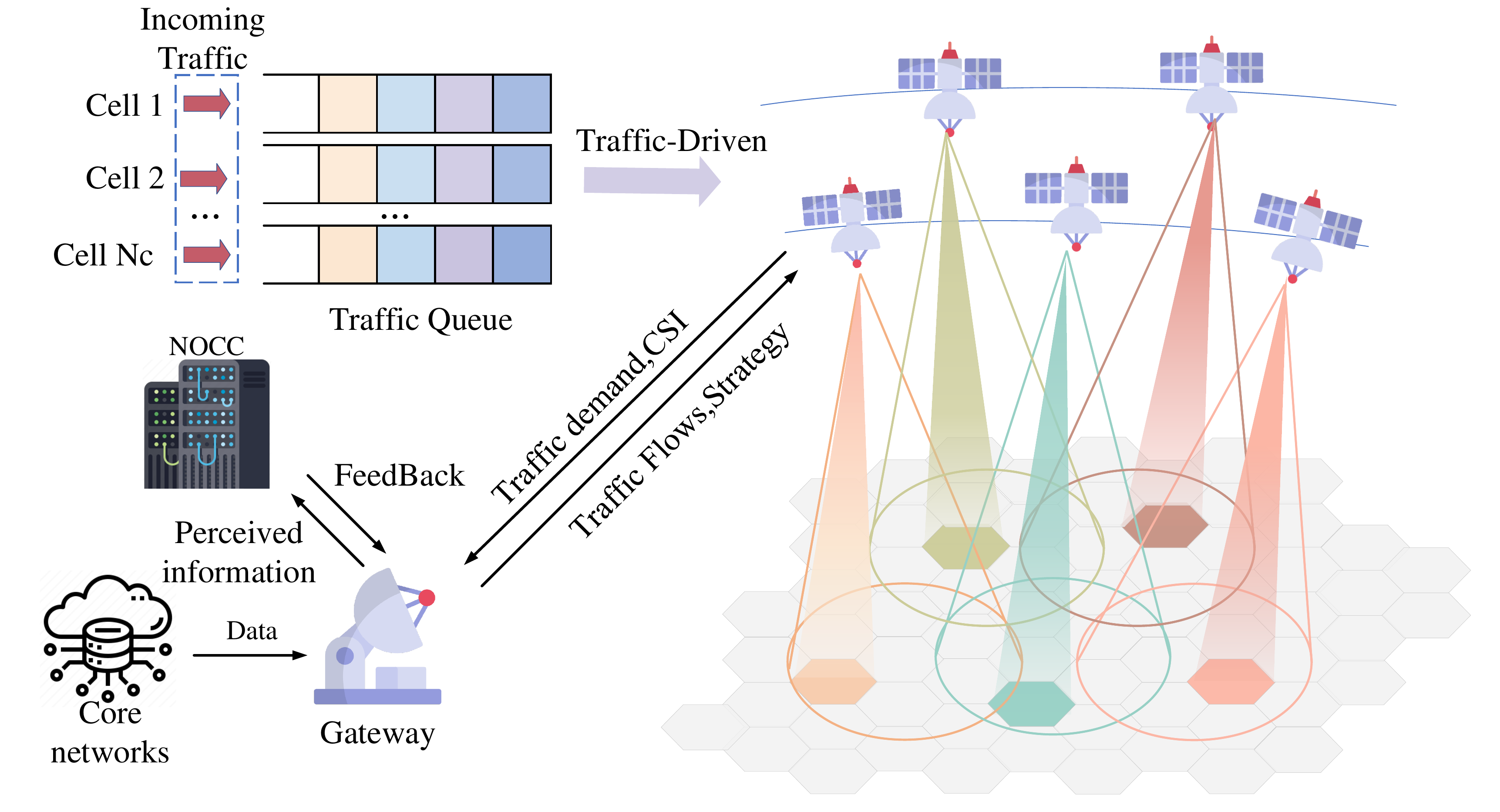}}
\caption{The forward link of the multi-NGSO beam hopping communication system.}
\label{fig:0}
\vspace{0em}
\end{figure}

This paper investigates the multi-NGSO satellite BH scenario, encompassing multiple satellites, ground cells, gateways, core networks, and the network operations control center (NOCC), as illustrated in Fig. \ref{fig:0}.  It is assumed that each NGSO satellite is equipped with a phased array antenna that can dynamically generate a limited number of high-gain spot beams and flexibly adjust their illumination directions. The ground is divided into multiple cells of equal size. The traffic volume in each cell fluctuates over time, and there are also significant variations in traffic volume among the different cells. Assuming a single spot beam can precisely cover the cell. Given that the forward link typically handles higher traffic volumes, this study focuses on the forward link of the multi-NGSO BH system. The ground gateway gathers real-time information from each satellite, including channel status, traffic demand, and terminal location, and transmits the data to the NOCC. The NOCC analyzes the data to determine the resource allocation strategy. Finally, the core network sends the necessary data to satellites via the gateway.

\begin{figure}[!htbp]
\vspace{0em}
\centerline{\includegraphics[width=0.5\textwidth]{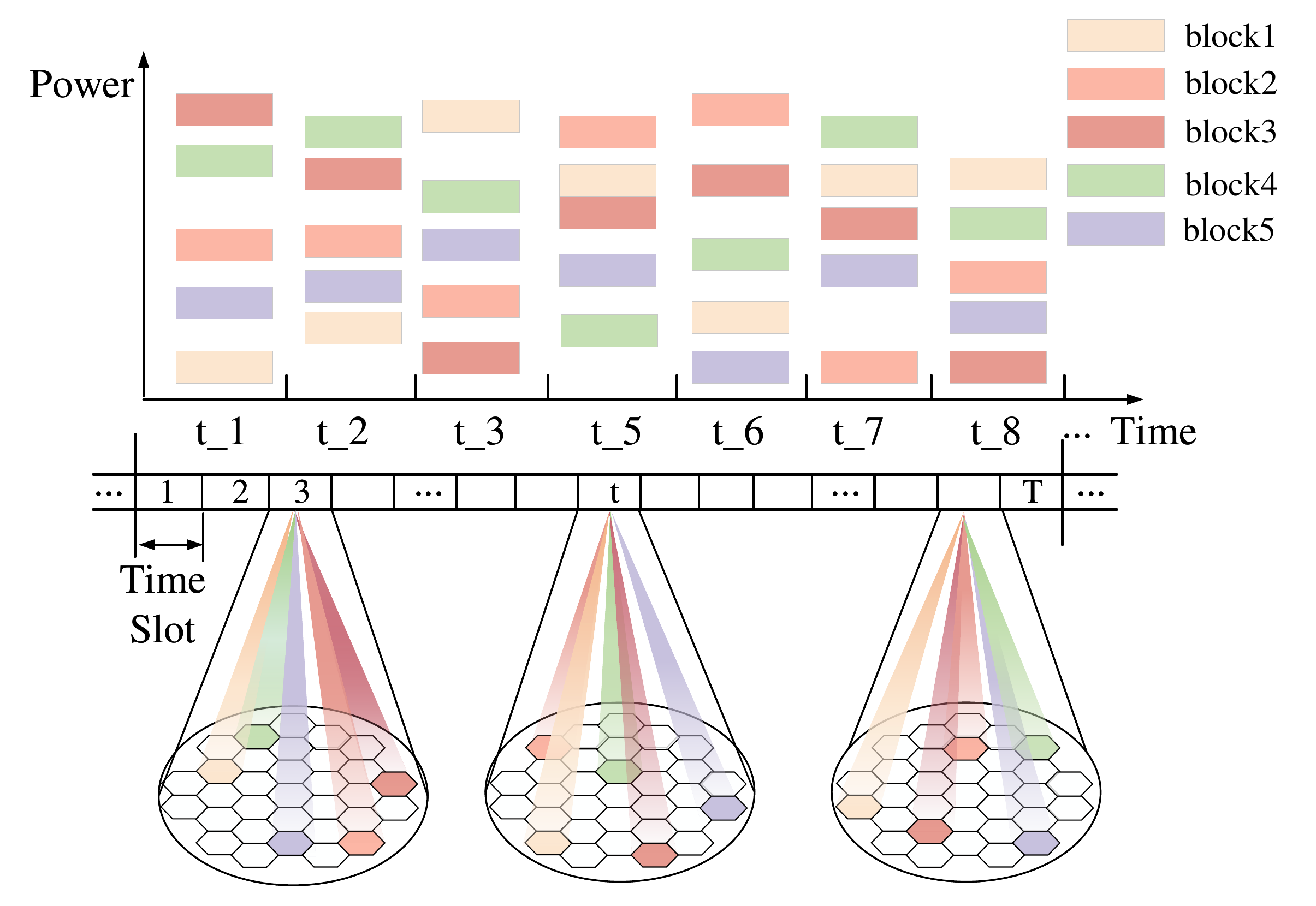}}
\caption{Beam hopping illumination patterns and power allocation.}
\label{fig:1}
\vspace{0em}
\end{figure}

Assuming there are $N_c$ virtual cells in the provided area, they can be represented by the set $\mathcal{N} = \{n | n = 1,2,\ldots, N_c \}$. There are $N_s$ satellites covering these cells, which can be represented by the set $\mathcal{I} = \{i | i = 1,2,\ldots, N_s \}$. The subset of virtual cells covered by each satellite $i$ is defined as $V_i$. It is important to note that some virtual cells are covered by multiple satellites. Therefore, the union of all subsets $V_i$ across the $N_s$ satellites will collectively cover all $N_c$ virtual cells in the area. This coverage method ensures that each virtual cell is covered by at least one satellite, achieving full coverage of the entire area. In the BH system, each satellite can generate up to $K$ beams and provide coverage for virtual cells through time division multiplexing. According to the current strategy, the satellite serves different cells at different times and allocates the corresponding power to each beam, as shown in Fig. \ref{fig:1}. The satellite illuminates different cells in different time slots, and the power allocated to each beam is also different. The different colors of the resource block represent different beams, and the figure is just an example. Not all beams are depicted.

To satisfy the traffic demand of the virtual cells, the satellites must determine the irradiation position of the beams (i.e., which $K$ virtual cells should be served at each time slot). 
Therefore, the BH pattern of the $i$-th satellite can be represented as 
\begin{equation}
    X_i=\left\{x_i^1, x_i^2, ...,x_i^t,...,x_i^T\right\},
\end{equation}
where $x_i^t$ denotes the BH pattern of the $i$-th satellite at time slot $t$. It can be expressed as $x_i^t=\left\{x_{i,1}^t, x_{i,2}^t,..., x_{i,n}^t,...,x_{i,N_c}^t\right\} $, where $x_{i,n}^t\in\{0,1\}$ indicates whether virtual cell $n$ is illuminated by satellite $i$ at time slot $t$. We denote $\mathcal{T} = \{t | t = 1,2,\ldots, T \}$ as the set of time slots.

\subsection{Communication Model}
In a multi-NGSO BH scenario, assuming each satellite can generate $K$ beams on the forward link. Each virtual cell is equipped with antennas that compensate for the Doppler shift caused by the satellite motion. Additionally, the environmental conditions are assumed to be ideal, with clear skies and no rain attenuation affecting signal quality. Based on these assumptions, the channel gain between the $i$-th satellite and the $n$-th cell can be further expressed as \cite{r35}
\begin{equation}
    h_{i,n}^t = G_t(\theta_{i,n}^t) \cdot L_{i,n}^t \cdot G_r(\theta_{i,n}^t) ,
\end{equation}
where $G_t(\theta_{i,n}^t)$ and $G_r(\theta_{i,n}^t)$ represent the transmit antenna gain of satellite $i$ to cell $n$ and the receive antenna gain of cell $n$ from satellite $i$, respectively. $L_{i,n}^t$ denotes the path loss between satellite $i$ and ground user $n$ at time slot $t$, expressed as
\begin{equation}
    L_{i,n}^t = \frac{1}{{4\pi\frac{{l_{i,n}^t}^2}{\lambda}}},
\end{equation}
where $l_{i,n}^t$ is the distance between satellite $i$ and virtual cell $n$ at time slot $t$. $\lambda$ is the wavelength of the transmitted signal.

The antenna radiation pattern for the multi-beam antenna is based on the model in 3rd Generation Partnership Project (3GPP) TR 38.811 \cite{r36}, which can be described as
\begin{equation}
    G_t(\theta) = 
\begin{cases} 
1, & \theta = 0^\circ \\
\frac{4J_1(2\pi a \sin(\theta/\lambda))^2}{2\pi a \sin(\theta/\lambda)}, & 0^\circ < |\theta| \leq 90^\circ
\end{cases},
\end{equation}
where $\lambda$ represents the wavelength, and $a$ denotes the radius of the antenna's circular aperture. The function $J_1(\cdot)$ is the first-order Bessel function of the first kind. The off-axis angle, measured in degrees, is denoted by $\theta$. At the peak orientation, the antenna exhibits its maximum gain $G_m$. The user's receiving antenna is considered omnidirectional, characterized by a uniform gain in all directions, quantified as $0$ dBi.

Given the limited transmission power of satellites, effective power allocation is essential for optimizing the overall power efficiency of satellite systems. The total power available for transmission is denoted by $P_{tot}$. In this paper, $p_{i,n}^t$ represents the power allocated by satellite $i$ to the illuminate cell $n$ at time slot $t$. This paper adopts full-frequency multiplexing to optimize spectrum resources and improve utilization, which also means that the problem of CCI between beams cannot be ignored. The SINR of the cell $n$ can be expressed as

\begin{equation}
\begin{split}
    \text{SINR}_t^n = 
    \frac{p_{i,n}^t h_{i,n}^t}{k_B T_{rx} B +  \displaystyle\sum_{j=i} \displaystyle\sum_{\kappa \neq n} p_{j,\kappa}^t h_{j,\kappa}^t 
    + \displaystyle\sum_{j \neq i} \displaystyle\sum_{\kappa \neq n} p_{j,\kappa}^t  h_{j,\kappa}^t},
\end{split}
\end{equation}
where $k_B$ denotes the Boltzmann constant and $T_{rx}$ refers to the receiver noise temperature. Satellite interference consists of two parts. The first is intra-satellite interference, which comes from beams of the same satellite. The second is inter-satellite interference, which comes from beams of other satellites. Then, the rate of cell $n$ in time slot $t$ can be expressed as
\begin{equation}
    R_t^n = Blog_2(1+\rm{SINR}_t^n),
\end{equation}

\subsection{Queue Model}
We consider that ground users are unevenly distributed and vary over time and that the traffic demand for each virtual cell differs. Given the limited communication and storage resources, assume that there are $N_c$ queues for storing incoming traffic in each cell. To account for the timeliness of the traffic, we assume each queue stores only the most recent $T_{ttl}$ time slots of traffic data, where $T_{ttl}$ represents the time-to-live of the data. Traffic in the queue that exceeds the $T_{ttl}$ will be discarded. Furthermore, newly incoming traffic must be dropped if it exceeds the queue's storage capacity. Traffic is processed in the queue on a first-come, first-served (FIFO) basis. All satellites access a common database for synchronization.

The backlog in the queue (or the size of the queue) in a time slot $t$ is denoted by $Q_t = \{Q_t^n | n \in \mathcal{N}\}$, where $Q_t^n$ represents the backlog of queue $n$ at time slot $t$. The queue evolves according to the following equation
\begin{equation}
    Q_{t+1}^n = (Q_t^n-L_t^n +A_t^n-D_t^n)^+, \forall{n}.
\end{equation}
where $(x)^+ = max\left\{x,0\right\}$. $L_t^n$ represents the number of data packets processed in queue $n$ during time slot $t$, $A_t^n$ is the amount of the new incoming data at queue $n$, and $D_t^n$ is the amount of date discarded due to timeout. We consider the early arrival late departure model for the queue.

As mentioned, data are discarded if they exceed the time to live or the storage capacity of the queue. This implies that all queues in this system are finite and strongly stable \cite{r43}.

We denote $\kappa_t^n \in {0,1}$ as the server decision variable. Specifically, $\kappa_t^n = 1$ indicates that the cell $n$ is illuminated (i.e., served) at time slot $t$. The number of packets processed in time slot $t$ is calculated as
\begin{equation}
    L_t^n = min\left\{\frac{\kappa_t^n R_t^n T_{slot}}{o}, Q_t^n\right\}.
\end{equation}
where $o$ represents the size of the data packet and $T_{slot}$ is the duration of each time slot.
The number of packets arriving in slot $t$ is $A_t^n$ , where $A_t^n$ follows the Poisson distribution with arrival rate $\rho_t^n$.

\section{Problem Formulation}

This paper mainly studies the traffic-driven beam hopping optimization problem. According to the dynamic changes in users' traffic requirements, the beam-hopping and resource allocation strategies can be flexibly adjusted to meet traffic demands in different spatiotemporal scenarios. We focus on the throughput and delay performance of each time slot because resource allocation and user requirements in satellite networks are highly dynamic and complex. Time-by-time optimization helps maximize the use of resources and meet current demands at each instantaneous stage, improving the quality of service. While the system's long-term performance is represented as an overall effect across all time slots, it essentially depends on the specific decisions and performance of each time slot. In other words, optimizing each time slot determines the resource utilization and service quality of the system under instantaneous conditions, and the accumulation of these instantaneous performances ultimately forms the system's long-term performance. Therefore, by optimizing the performance of each time slot, the overall operating performance and long-term performance can be effectively improved.

The amount of data transmitted by cell $n$ at the time slot $t$ can be expressed as
\begin{equation}
    T_t^n = min \{\kappa_t^n R_t^n T_{slot},Q_t^n o\},
\end{equation}
The long-term network throughput can be expressed as
\begin{equation}
    \Upsilon = \sum_{t=1}^T \sum_{n=1}^{N_c} T_t^n.
\end{equation}


Since each packet in queue $n$ has a different waiting time, the backlog $Q_t^n$ at time $t$ can be decomposed into $\{\phi_{t,1}^n, \phi_{t,2}^n,..., \phi_{t,l}^n,...,\phi_{t, T_{ttl}}^n\}$. The total backlog is the sum of packets in all waiting time categories $\sum_{l=1}^{T_{ttl}} \phi_{t,l}^n = Q_t^n$. 
Here, $\phi_{t,l}^n$ represents the number of data packets in queue $n$ that have been waiting for $l$ time slots at time slot $t$.

The number of packets in each waiting time category, $\phi_{t,l}^n$, is determined by the number of packets processed in the previous time slot and the remaining backlog. It can be expressed as 
\begin{equation}
    \phi_{t,l}^n = 
\begin{cases} 
0, & L_{t-1}^n > \sum_{l=1}^{l-1} \phi_{t-1,l}^n\\
\sum_{l=1}^{l-1} \phi_{t-1,l}^n - L_{t-1}^n, &  L_{t-1}^n < \sum_{l=1}^{l-1} \phi_{t-1,l}^n
\end{cases},
\end{equation}

Next, we consider the queuing delay of the traffic in each cell. Let $\tau_n^t$ denotes the average queue delay for each packet in cell $n$ at time slot $t$, which can be calculated by \cite{r42}
\begin{equation}
    \tau_t^n = \frac{\sum_{l=1}^{T_{ttl}}l \phi_{t,l}^n}{\sum_{l=1}^{T_{ttl}} \phi_{t,l}^n},
\end{equation}
where $\phi_{t,l}^n$ represents the number of data packets in queue $n$ that have been waiting for $l$ time slots at time slot $t$. It is worth noting that we have fully considered the coupling between multiple queues and included its effect on the delay in the calculation. $\phi_{t,l}^n$ is the delay under this coupling effect.

Due to the significant impact of dynamic traffic and resource allocation strategies on queuing delay, this paper primarily focuses on its optimization. 
We define the long-term cumulative average delay (LTCAD) as \cite{r42}
\begin{equation}
    \Gamma= \frac{1}{T N_c}\sum_{t}\sum_{n} \tau_t^n.
\end{equation}

The objective is to maximize the total network throughput while minimizing LTCAD simultaneously. To achieve this, the overall utility function $G$ combines the long-term network throughput and LTCAD into a single optimization target. A weighting factor $\alpha$ is introduced to balance the trade-off between these two objectives. It can be expressed as
\begin{equation}
    G = \alpha \frac{\Upsilon}{\Upsilon_{norm}} - (1-\alpha) \frac{\Gamma}{\Gamma_{norm}},
\end{equation}
where $\Upsilon_{norm}$ and $\Gamma_{norm}$ are normalization constants.

Therefore, we cast the traffic driven BH scheduling problem as an optimization problem with the goal of maximizing the total utility function $G$, expressed as
\begin{equation}
\begin{aligned}
    \max_{\mathbf{x}, \mathbf{p}} \quad&
    G \\
    s.t. \quad
    & C1: \sum_{n=1}^{N_c} x_{i,n}^t \le K, \quad x_{i,n}^t \in \{0,1\}, \quad \forall i,n,t, \\
    & C2: \sum_{n=1}^{N_c} p_{i,n}^t \le P_{\text{tot}}, \quad \forall t, \\
    & C3: P_{\text{min}} \le p_{i,n}^t \le P_{\text{max}}, \quad \forall i,n,t .\\
\end{aligned}
\end{equation}
The decision variables representing the beam irradiation direction and power allocation for each beam at each time slot can be denoted as $\mathbf{x}=\left\{x_{i,n}^t| i\in \mathcal{I}, n\in \mathcal{N}, t\in \mathcal{T} \right\}$ and $\mathbf{p}=\left\{p_{i,n}^t| i\in \mathcal{I}, n\in \mathcal{N}, t\in \mathcal{T} \right\}$ respectively. $C1$ indicates that each satellite generates at most $K$ beams simultaneously. $C2$ ensures that the total power of all satellite beams does not exceed the total satellite power $P_{tot}$. $C3$ limits the power of each beam to between $P_{min}$ and $P_{max}$.

Observations indicate that problem (15) presents non-convex and nonlinear challenges, primarily due to the nonlinearity of SINR described in Eq. (5). The introduction of binary variables $p_{i,n}^t$ and $x_{i,n}^t$ further exacerbates this complexity and makes the optimization problem NP-hard. Therefore, obtaining an optimal solution in polynomial time is infeasible. The objective function in problem (15) has long-term accumulation characteristics, as illustrated by Eq. (10) and Eq. (13), highlighting the need for vertical optimization methods. This characteristic reframes the optimization challenge as a sequential decision-making problem, for which DRL techniques offer a promising solution. The details of exploring this complex optimization environment using DRL methods will be presented in detail in the subsequent sections.

\section{DRL for Beam Hopping and Power Allocation}

Due to its flexibility, efficiency, and adaptability, DRL offers significant advantages in solving joint optimization problems and effectively addressing complex non-convex challenges. Unlike traditional optimization methods that rely on problem decomposition and convex relaxation, DRL processes high-dimensional variables directly through gradient descent and deep neural networks. In addition, DRL can learn effective long-term strategies through continuous interaction with the environment. Therefore, we have designed a DRL-based optimization framework to jointly optimize beam illumination patterns and power allocation, enabling adaptive optimal decision-making in high-uncertainty and dynamics scenarios.

In DRL, decision-making problems are typically represented as MDPs. An MDP can be described by a five-tuple $(S, A, P, R, \gamma)$, where $S$ represents the set of states in the environment, $A$ represents the set of actions, $P$ represents the state transition probabilities, $R$ represents the reward function, and $\gamma \in [0, 1]$ is the discount factor, which adjusts the relative weight of immediate rewards versus future rewards. The discount factor $\gamma$ affects the impact of future rewards on current decisions. At time step $t$, $a_t$ represents the action chosen by the agent, and $s_t$ represents the observed state. After taking action $a_t$, the agent receives a reward $r_t$ as performance feedback, reflecting the effectiveness of action $a_t$. The agent's goal is to maximize the cumulative discounted reward $R_t = \sum_{\tau=0}^{\infty} \gamma^\tau r_{t+\tau}$, where the cumulative reward is the weighted sum of all future rewards from the time step $t$ onward. However, accurately identifying state transition probabilities is challenging in practical applications, making model-free DRL algorithms particularly important. These algorithms do not rely on prior knowledge or detailed mathematical models, but instead learn the best strategy through continuous interaction with the environment. The agent continuously experiments and receives feedback to adjust its strategy to maximize the preset objective function. This approach enables model-free DRL algorithms to optimize decision-making without detailed environmental information. Through interactions with the environment, the agent learns to choose actions that lead to the highest cumulative discounted rewards. This process helps the agent optimize its decision-making strategy to perform best in a dynamic environment \cite{r38}.

\subsection{MDP Formulation}

The amount of data $Q_t^n$ at time $t$ depends on the current time $t$ and the previous time $t-1$, and its variation over time follows the characteristics of a discrete-time MDP. This implies that the current amount of data depends not only on the previous state but also on the current input and processing conditions. As a result, the system's state transitions and reward mechanisms can be effectively described and optimized using a discrete-time MDP model.

Existing algorithms typically break down the joint optimization problem into multiple sub-problems, such as beam illumination pattern and resource allocation, when addressing the BH problem. Specifically, the algorithm initially determines the illumination position of each beam based on a predefined criterion, such as maximizing coverage or minimizing interference. After establishing the beam illumination pattern, the algorithm allocates resources to optimize system performance. The advantage of this approach is that it simplifies the complex multi-satellite BH scheduling problem by breaking it down into two more manageable sub-problems, thus reducing computational complexity and making the problem easier to handle. However, this decomposition method also has certain limitations. Beam illumination position and resource allocation are interdependent. In total frequency reuse, it is crucial to consider both factors simultaneously. To address these challenges, we model the problem in Eq. (15) as an MDP and develop a DRL-based optimization framework to optimize BH patterns and power allocation jointly. In this process, the key elements of the MDP—state, action, and reward function—are critical in RL, as they directly influence the agent’s learning effectiveness and the attainment of optimization goals. Therefore, we first define these three essential elements to ensure that the optimization framework effectively addresses the problem and enhances system performance.

\textit{1) State:}
In multi-NGSO BH scheduling, decisions are influenced by traffic demand and the state of the network. Thus, state definition needs to incorporate these key characteristics. We define the state as
\begin{equation}
    s_t = \left\{Q_t,H_t\right\}.
\end{equation}
Where $Q_t$ represents the cumulative traffic data in the queue at time slot $t$, it can be described as $Q_t=\left\{Q_t^1,Q_t^2,...,Q_t^{N_c}\right\}$. $H_t$ represents the channel gain matrix.
\begin{equation}
    H_t = \begin{bmatrix}
    h_{1,1}^t & h_{1,2}^t & \cdots & h_{1,N_c}^t\\
    h_{2,1}^t & h_{2,2}^t & \cdots & h_{2,N_c}^t\\
    \vdots & \vdots & \ddots & \vdots\\
    h_{N_s,1}^t & h_{N_s,2}^t & \cdots & h_{N_s,N_c}^t
    \end{bmatrix},
\end{equation}
where $h_{i,n}^t$ represents the channel gain from satellite $i$ to cell $n$ at time slot $t$.

\textit{2) Action:} In multi-NGSO BH communication systems, the definition of action is crucial because it directly affects the performance and efficiency of the system. Based on the decision variables defined in (15), the agent must decide the direction of illumination of the beam and the allocation of power in each time interval. Therefore, we define an action as

\begin{equation}
    a_t = (w_t,v_t),
\end{equation}
where $w_t$ denotes the BH pattern at time slot $t$, which can be expressed as 

\begin{equation}
    w_t = \begin{bmatrix}
    w_{1,1}^t & w_{1,2}^t & \cdots & w_{1,K}^t\\
    w_{2,1}^t & w_{2,2}^t & \cdots & w_{2,K}^t\\
    \vdots & \vdots & \ddots & \vdots\\
    w_{N_s,1}^t & w_{N_s,2}^t & \cdots & w_{N_s,K}^t
    \end{bmatrix}.
\end{equation}
$v_t$ represents the power allocation of the beams at time slot $t$, which can be expressed as 

\begin{equation}
    v_t = \begin{bmatrix}
    v_{1,1}^t & v_{1,2}^t & \cdots & v_{1,K}^t\\
    v_{2,1}^t & v_{2,2}^t & \cdots & v_{2,K}^t\\
    \vdots & \vdots & \ddots & \vdots\\
    v_{N_s,1}^t & v_{N_s,2}^t & \cdots & v_{N_s,K}^t
    \end{bmatrix}.
\end{equation}

\textit{3) Reward:} 
The reward function evaluates the immediate impact of taking action $a_t$ in state $s_t$. While our overall goal is to optimize the system's long-term performance, this optimization is achieved through specific action decisions for each time slot. Therefore, the immediate reward for each time slot includes at least the throughput performance of that time slot and the queuing delay under queue coupling. We define the reward function as
\begin{equation}
    r_t=\alpha \frac{\Upsilon}{\Upsilon_{norm}} - (1-\alpha) \frac{\Gamma}{\Gamma_{norm}} - penalty_b - penalty_p,
\end{equation}
where the first term $\Upsilon = \sum\limits_{n=1}^{N_c} T_t^n$ denotes the throughput at time slot $t$ and the second one $\Gamma = \frac{1}{N_c} \sum\limits_{n=1}^{N_c} \tau_t^n$ represents the queuing delay at time slot $t$. 
The $penalty_b$ represents the penalty for multiple beams that cover the same unit, with its intensity proportional to the number of beams involved. The $penalty_p$ accounts for exceeding the satellite's total power limit, with intensity proportional to how much the beam power exceeds this limit. Together, these penalties reflect beam concentration and power excess, respectively. After defining the three key elements, we will introduce the beam illumination and power allocation algorithm based on PPO.

\subsection{Algorithm Design}
\begin{figure*}[!htbp]
\vspace{0em}
\centerline{\includegraphics[width=0.85\textwidth]{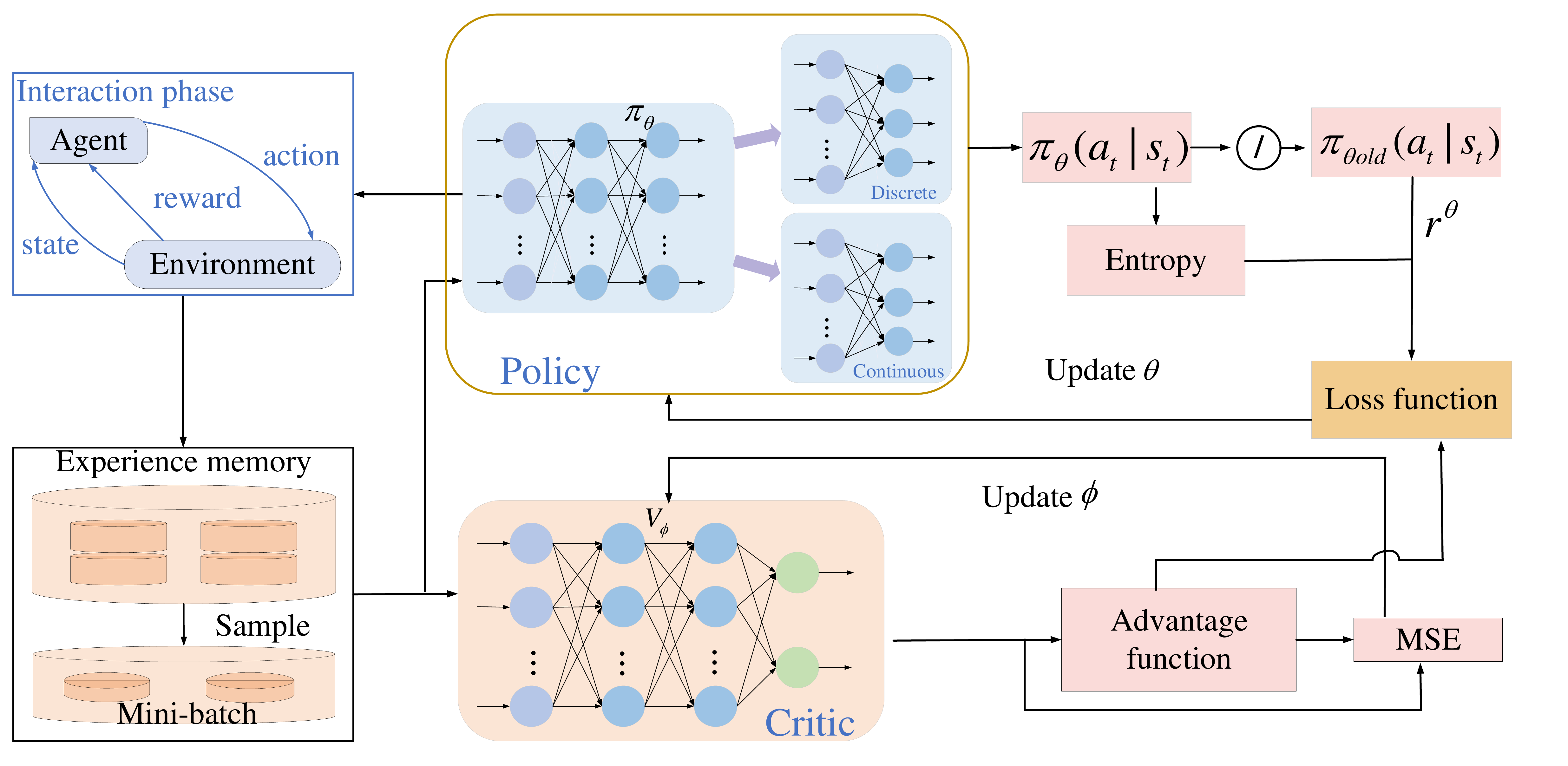}}
\caption{PPO framework for the proposed optimization problem}
\label{fig:2}
\vspace{0em}
\end{figure*}

The PPO algorithm is widely utilized in DRL due to its simplicity and robustness. It integrates the benefits of both value function and policy gradient methods. Unlike traditional policy gradient (PG) algorithms, PPO introduces a probability ratio clipping function and importance sampling to constrain the magnitude of policy updates. This approach improves sample efficiency and reduces the sensitivity of policy updates to changes in trajectories.

The policy network $\pi_\theta$, with weights $\theta$, is primarily responsible for interacting with the environment and determining the actions to take. In contrast, the critic network $V_\phi$, with weights $\phi$, evaluates the performance of the policy network by estimating the value of states. Specifically, the policy network interacts continuously with the environment, generating experience data. These data are stored in the memory pool in the form of $(s_t, a_t, r_t, s_{t+1}, \pi_{\theta_{old}(a_t|s_t))}$. The action $a_t$ consists of discrete and continuous components, representing the beam illumination pattern and power distribution, respectively. The experience data collected by the policy network is then used to update the weights of both the policy and value networks multiple times. This iterative process allows the policy network to improve its decision-making strategy. In contrast, the critic network provides feedback by evaluating the effectiveness of the policy network's actions based on the state value function.

This paper aims to optimize BH illumination patterns and power allocation to maximize system throughput and minimize queuing delay. Addressing the mixed action space comprising discrete and continuous components, traditional algorithms often employ discretization methods to convert the continuous action space into a discrete form. However, such methods risk the loss of critical information.

To address this limitation, this work employs two independent and parallel policy networks to separately manage the discrete and continuous components of the action space, as illustrated in Fig. \ref{fig:2}. These networks share the initial layers of a base network, which are dedicated to extracting shared features. Beyond the shared layers, the discrete policy network produces a probability distribution over beam illumination patterns, selecting the final discrete action through sampling. Concurrently, the continuous policy network outputs the mean and standard deviation of the power distribution, determining the final continuous action via sampling from a Gaussian distribution defined by these parameters.

This dual-network approach enhances the accuracy and efficiency of action selection by separately processing discrete and continuous actions while leveraging shared feature extraction. Thus, critical information is preserved, and overall system performance is improved.

After a certain training period, the agent samples a batch of experience data from the replay buffer. These experience data include the probability distribution of the old policy, represented as $\pi_{\theta_{old}}$, which contains the probability distribution of discrete policy $\pi_{\theta_{d_{old}}}$ and the probability density function of continuous policy $\pi_{\theta_{c_{old}}}$. Subsequently, the agent updates the policy network weights $\theta$ by optimizing the clipped objective function and the value network weights $\phi$ by minimizing the mean squared error.

Before introducing the loss function $L(\theta)$ for policy network, we first give the definition of the advantage function at time step $t$ as 
\begin{equation}
   \hat{A}(t) = \delta_t + (\gamma \lambda) \delta_{t+1}+\cdot \cdot \cdot+(\gamma \lambda)^{T-t+1} \delta_{T-1},
\end{equation}
where $\delta_t = r_t+\gamma V(s_{t+1})-V(s_t)$. $\delta_t$ is the temporal difference error at time step $t$, $\gamma$ is the discount factor, and $\lambda$ is the GAE-Lambda parameter, which is used to balance bias and variance.

PPO adopts a clipped surrogate advantage loss defined as 
\begin{equation}
    L(\theta) = \mathbb{E}\left[\min \left( r^{\theta}(t)  \hat{A}(t), \text{clip}(\cdot) \hat{A}(t) \right) \right].
\end{equation}
The probability ratio is defined as $r^{\theta}(t)=\frac{\pi_{\theta}(a_t|s_t)}{\pi_{\theta_{old}}(a_t|s_t)}$.
The clip operation $\text{clip}(\cdot)$ is short for $\text{clip}\left(r^{\theta}(t), 1-\epsilon, 1+\epsilon\right)$ to restrict the probability ratio into the interval $[1-\epsilon, 1+\epsilon]$ , which can prevent the large changes in policy updates. 

For discrete policy networks, the clipped surrogate advantage loss is defined as
\begin{equation}
    L_d(\theta) = \mathbb{E}\left[\min \left( r^{\theta}_d(t)  \hat{A}(t), \text{clip}\left(r^{\theta}_d(t), 1-\epsilon, 1+\epsilon\right) \hat{A}(t) \right) \right],
\end{equation}
where $r^{\theta}_d(t)=\frac{\pi_{\theta_d}(a_t|s_t)}{\pi_{\theta_{dold}}(a_t|s_t)}$. $\pi_{\theta_d}(a_t|s_t)$ and $\pi_{\theta_{dold}}(a_t|s_t)$ represent the discrete action probability of the current policy and the discrete action probability of the old policy, respectively. For continuous policy networks, the clipped surrogate advantage loss is defined as
\begin{equation}
    L_c(\theta) = \mathbb{E}\left[\min \left( r^{\theta}_c(t)  \hat{A}(t), \text{clip}\left(r^{\theta}_c(t), 1-\epsilon, 1+\epsilon\right) \hat{A}(t) \right) \right],
\end{equation}
where $r^{\theta}_c(t)=\frac{\pi_{\theta_c}(a_t|s_t)}{\pi_{\theta_{cold}}(a_t|s_t)}$. $\pi_{\theta_c}(a_t|s_t)$ and $\pi_{\theta_{cold}}(a_t|s_t)$ represent the continuous action probability density of the current policy and the continuous action probability density of the old policy respectively.
Due to the shared base layers of the networks, so
\begin{equation}
    L(\theta) = L_d(\theta)+L_c(\theta).
\end{equation}

To incentivize the agent to explore more potential actions, we have added an entropy regularization method to the loss function of the behavior network. This approach encourages exploration and helps the agent steer clear of suboptimal scenarios. Thus, we have replaced the loss function of the behavior network with
\begin{equation}
    L(\theta) = L(\theta)+S_{\pi_\theta},
\end{equation}
where $S_{\pi_\theta}$ denotes the information entropy of the new policy, including the information entropy of the discrete strategy $S_{\pi_{d\theta}}$ and the information entropy of the continuous network strategy $S_{\pi_{c\theta}}$. 
\begin{equation}
    S_{\pi_\theta} = S_{\pi_{d\theta}}+S_{\pi_{c\theta}},
\end{equation}
where $S_{\pi_{d\theta}} = -\sum\limits_{\pi_{\theta_d}(a_t|s_t)}\pi_{\theta_d}(a_t|s_t)log\pi_{\theta_d}(a_t|s_t)$ and $S_{\pi_{c\theta}}=\frac{1}{2}+\frac{1}{2}log(2\pi\sigma^2)$. $\sigma^2$ is the variance.


The loss of the critic network is determined by computing the mean squared error between its output and the advantage function.
\begin{equation}
L(\phi)=\mathbb{E}\left[\frac{1}{N}\sum_{t=1}^N(\hat{A}(t)-V(s_t))^2 \right].
\end{equation}

Algorithm. \ref{alg.0} describes the overall process of the proposed PPO scheme in detail in the form of pseudo-code.

\begin{algorithm}
    \caption{Beam illumination patterns and power allocation algorithm based on PPO.}
    \label{alg.0}
    \begin{algorithmic}
    \STATE Initialize policy network $\pi_{\theta}$ and critic network $V_\phi$.
    \STATE Initialize replay memory $\mathcal{D}$, mini-batch size $\mathcal{B}$, the learning rate of policy and critic network, discount factor $\gamma$, GAE parameter $\lambda$, clip parameter $\epsilon$, number of episodes $\mathcal{EP}$, number of steps $\mathcal{ST}$, and multiple epochs update number $\mathcal{KE}$.
    
    
    \FOR{episode = 1 to $\mathcal{EP}$}
       \FOR{t = 1 to $\mathcal{ST}$}
            \STATE Get the channel observation $H_t$, the queue length observation $Q_t$, and set $s_t = \left\{Q_t,H_t\right\}$.
            \STATE Input $s_t$ to the policy network and output the probability distribution of beam illumination the mean and standard deviation for power allocation.  
            \STATE Generate beam illumination pattern by sampling from this distribution. Generate power allocation action from the Gaussian distribution defined by the mean and standard deviation.  
            \STATE Obtain the reward $r_t$ and the environment steps into the next state $s_{t+1}$ after execute the joint action $a_t$. 
            \STATE Store transition $\left\{s_t,a_t,r_t,s_{t+1},\pi_{\theta_{old}}\right\}$ into relay memory $\mathcal{D}$. 
        \ENDFOR
            \IF{replay memory $\mathcal{D}$ is full}
                \FOR{i=1 to $\mathcal{KE}$}
                    \STATE Randomly sample the stored transitions from relay memory $\mathcal{D}$.
                    \STATE Update policy network $\theta$ and critic network $\phi$.
                \ENDFOR
                \STATE Update replay memory $\mathcal{D}$.
            \ENDIF
    \ENDFOR
    \end{algorithmic}
\end{algorithm}

\textbf{Other tricks:} We implement several techniques in the PPO algorithm to enhance training stability and efficiency. First, we adjusted the value range through advantage and state normalization to improve data distribution. Second, we employed orthogonal initialization for the neural network model parameters, which helps to reduce the likelihood of gradient disappearance and explosion. These techniques contribute to more stable training.

\subsection{Computational Complexity}
The computational complexity of the proposed algorithm depends on the complexity of policy policy and critic networks. The input layer size of both neural networks is determined by the size of the state space, denoted as $I_s=N_s N_c + N_c$. The policy network has two output layers that share a common base layer. The sizes of these two output layers are determined by the action space. The discrete network's output layer size is denoted as $O_d^a=\sum_{i=1}^{N_s} K|V_i|$, and the continuous network's output layer size is denoted as $O_c^a=2KN_s$. while the critic network’s output layer size equals 1, i.e., $O_s^c=1$. We assume that the policy and critic networks have the same architecture, with $L$ layers and $n_l$ neurons in each layer $l$. Based on these assumptions, we can derive the computational complexity of the proposed algorithm as follows.

\begin{equation}
\begin{aligned}
\mathcal{O}(2 I_s n_1+\sum_{l=1}^{L-1}n_l n_{l+1} +n_L (O_d^a + O_c^a)+ \sum_{l=1}^{L-1}n_l n_{l+1}+n_L O_s^c ) \\=
\mathcal{O}(N_s N_c n_1+\sum_{l=1}^{L-1}n_l n_{l+1} +n_L (\sum_{i=1}^{N_s} K|V_i| + KN_s )).
\end{aligned}
\end{equation}

\section{Simulation Results and Performance Analysis}
\subsection{Simulation Configuration}
In this paper, a Ku-band forward BH system is simulated. Table. \ref{tab2} summarizes the simulation parameters, and some parameters refer to the 3GPP standard and DVB-S2X standard \cite{r8}. There are 5 satellites distributed in the area, and 161 cells are covered. The orbit height is set as 550 km. As shown in Fig. \ref{fig:0}, it should be noted that some cells are covered by multiple satellites. The total power of each satellite is $P_{tot}$ = 39 dBW, and the maximum number of beams for each satellite is equal to $8$. The total bandwidth is set as $500$ MHz, and the duration of the time slot is set to $2$ ms, which refers to DVB-S2X standard 
 \cite{r8}.

To comprehensively demonstrate the superiority of our algorithm, we compare the proposed algorithm with the following three schemes:

1) SAC \cite{r0}: The SAC algorithm is used to solve the multi-objective optimization problem. The SAC algorithm operates in a MDP, which consists of five networks that are able to learn the optimal scheduling strategies and resource allocation methods after multiple training episodes.

2) Throughput Priority Beam Hopping (TP-BH): In the TP-BH algorithm, each satellite selects the $K$ cells with the largest traffic data in each time slot for service. Specifically, the algorithm compares the queue length of each cell and selects the 8 cells with the longest queues for service. The satellite's transmission power is then evenly distributed among these 8 cells.

3) Delay Priority Beam Hopping (DP-BH): In the DP-BH algorithm, each satellite selects the $K$ cells with the highest average delay in each time slot for service. The specific implementation method involves comparing the queue delays in each cell, selecting the 8 cells with the highest delays, and evenly distributing the satellite's transmission power among these 8 cells to minimize the task delays in these cells as much as possible, thereby improving the overall system performance.

4) User Service Weight Gain Priority Beam Hopping (USWGP-BH): Consistent with the method in \cite{r24}, this algorithm selects cells for service based on user service weight in each time slot. The user service weight is determined by the current traffic demand and delay of the cell. The algorithm calculates the service weight of each cell and selects the cells with the highest weights for service to optimize user experience and system efficiency.

\begin{table}[!htbp]
\vspace{-1.0em}
\caption{Main Simulation parameters}
\begin{center}
\begin{tabular}{l|l}
\hline
Parameter & value\\
\hline
Satellite orbit altitude $H$ (km) & 550\\
Ku-band frequency $f_c$ (GHz) &  12.4\\
Number of satellites $N_s$  & 5\\
Number of cells covered by all satellites $N_c$& 161\\
Radius of cell $R$  (km)& 14\\
The maximum beams for each satellite $K$ & 8\\
System bandwidth $B$ (MHz)& 100\\
Total satellite power $P_{tot}$ (dBW)& 39\\
The radius of the antennas aperture $a$ (m) & 0.15\\
3dB beamwidth of satellite beams & $3^o$\\
Maximum transmit antenna gain $G_m$ (dBi)& 35.9\\
Terminal receiving antenna gain $G_r$ (dBi)& 0\\
Noise temperature $T_{rx}$ (k) &290\\
Time slot duration $T_{slot}$ (ms) & 2\\

Replay memory $\mathcal{D}$  & 2048\\
Mini-batch sise $\mathcal{B}$  & 64\\
The learning rate of policy network & 3e-4\\
The learning rate of critic network & 3e-4\\
Discount factor $\gamma$ & 0.99\\
GAE parameter $\lambda$ & 0.95\\
Clip parameter $\epsilon$ & 0.2\\
Number of episodes $\mathcal{EP}$ & 10000\\
Number of steps $\mathcal{ST}$ & 200 \\
Optimizer & Adam\\
Activate function & tanh\\
$penalty_b$ & 0.005\\
$penalty_p$ & 0.005\\

\hline
\end{tabular}
\label{tab2}
\end{center}
\vspace{-0.5em}
\end{table}

\subsection{Result Analysis}

\begin{figure}[!htbp]
\vspace{-0.5em}
\centerline{\includegraphics[width=0.40\textwidth]{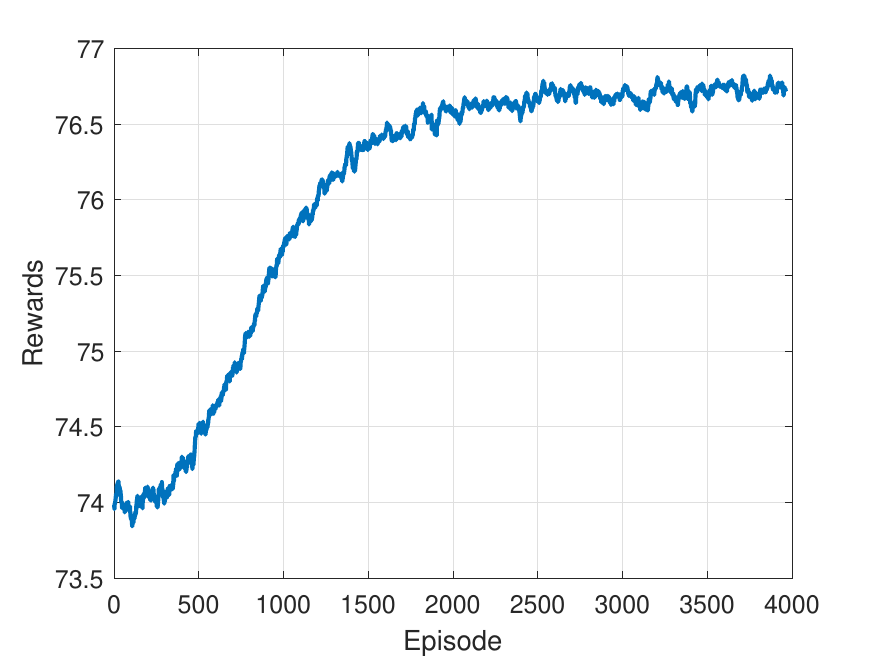}}
\caption{The convergence of reward.}
\label{fig:label}
\vspace{-0.65em}
\end{figure}

\begin{figure}[!htbp]
\vspace{-0.5em}
\centerline{\includegraphics[width=0.40\textwidth]{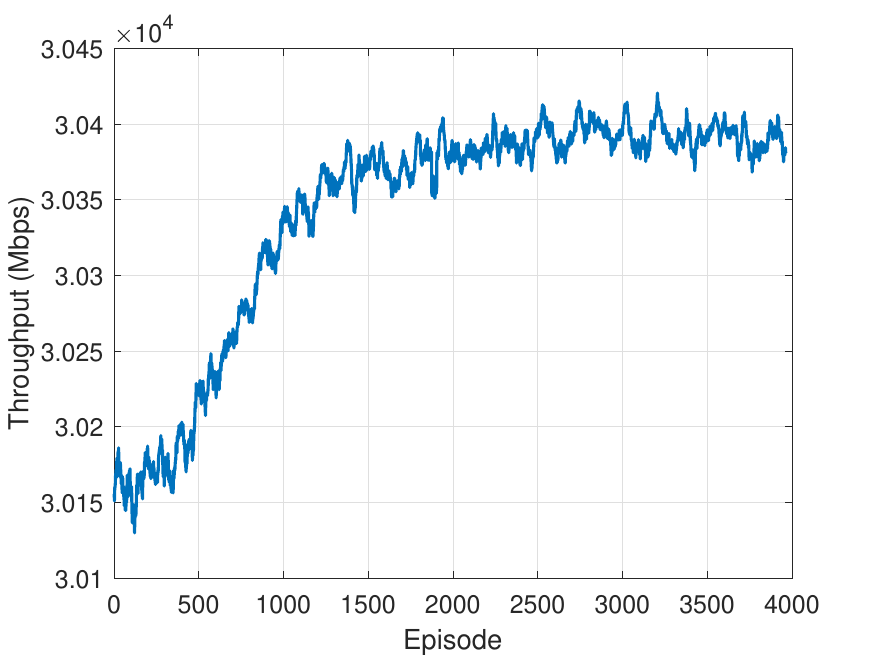}}
\caption{The convergence of throughput.}
\label{fig:labe2}
\vspace{-0.65em}
\end{figure}

\begin{figure}[!htbp]
\vspace{-0.5em}
\centerline{\includegraphics[width=0.40\textwidth]{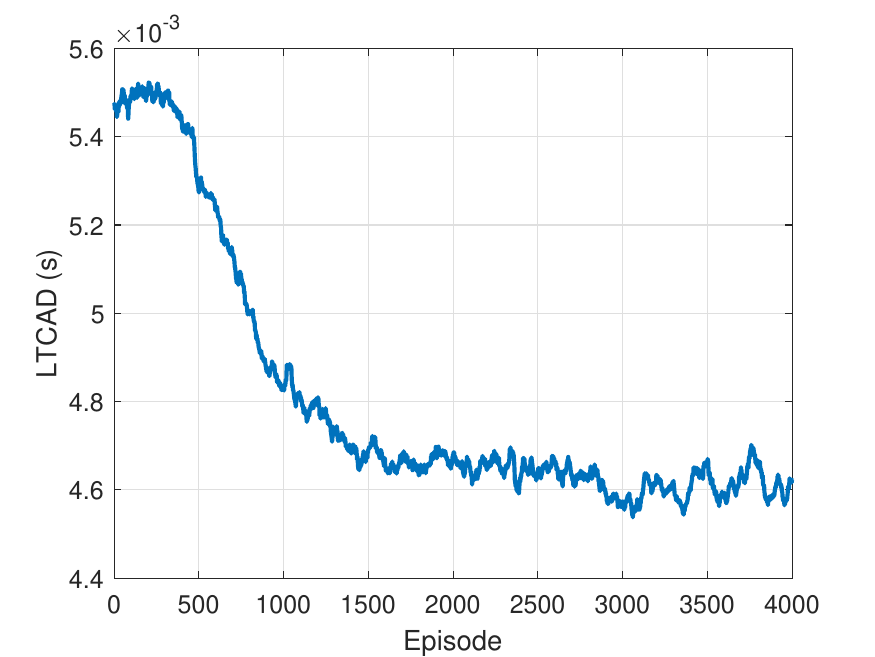}}
\caption{The convergence of LTCAD.}
\label{fig:labe3}
\vspace{-0.65em}
\end{figure}

To demonstrate the convergence of the proposed PPO-based algorithm, we illustrates the changes in episode reward, episode throughput, and episode latency during training. 
In Fig. \ref{fig:label}, the reward consistently increases with the number of training episodes, stabilizing after approximately 2500 episodes, indicating that the algorithm has reached a state of convergence. Fig. \ref{fig:labe2} shows that episode throughput experiences fluctuations during the early stages of training but gradually stabilizes after approximately 2500 episodes. This indicates that the algorithm can consistently optimize network throughput. In Fig. \ref{fig:labe3}, episode latency gradually decreases with an increase in training episodes, converging to 0.0046 after approximately 2500 episodes. The proposed algorithm effectively reduces network latency. These results confirm the convergence of the PPO-based algorithm under the parameter settings outlined in Table. \ref{tab2}.

\begin{figure}[!htbp]
\vspace{-0.5em}
\centerline{\includegraphics[width=0.40\textwidth]{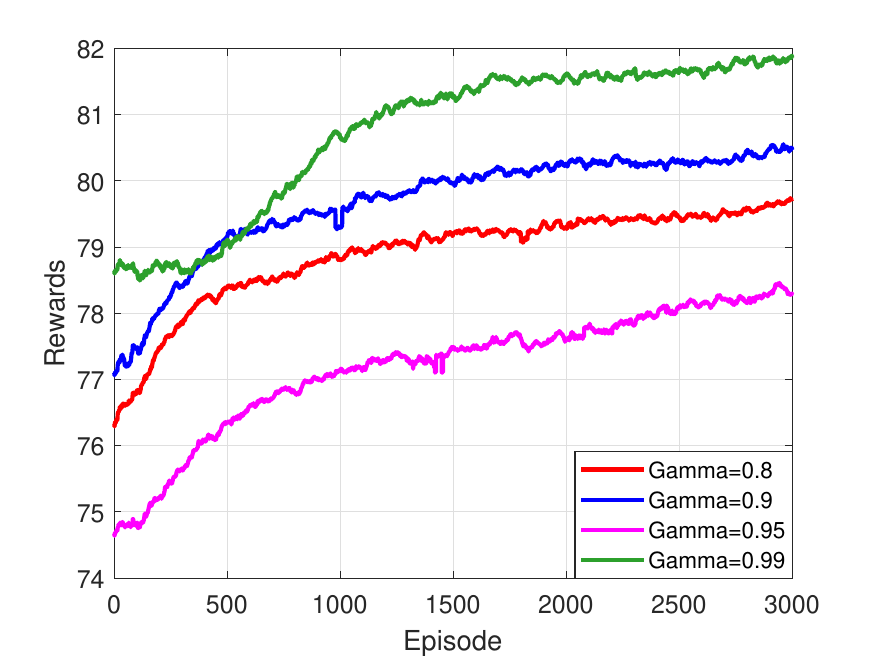}}
\caption{The comparison of convergence performance of different discount factors $\gamma$.}
\label{fig:labe4}
\vspace{-0.65em}
\end{figure}

\begin{figure}[!htbp]
\vspace{-0.5em}
\centerline{\includegraphics[width=0.40\textwidth]{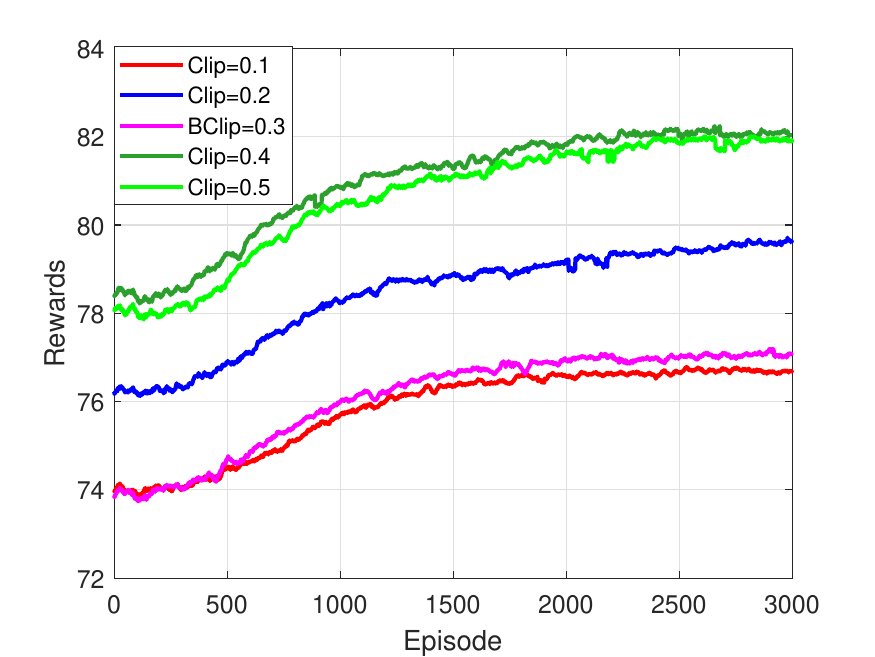}}
\caption{The comparison of convergence performance of different clip parameters $\epsilon$}
\label{fig:labe5}
\vspace{-0.65em}
\end{figure}

\begin{figure}[!htbp]
\vspace{-0.5em}
\centerline{\includegraphics[width=0.40\textwidth]{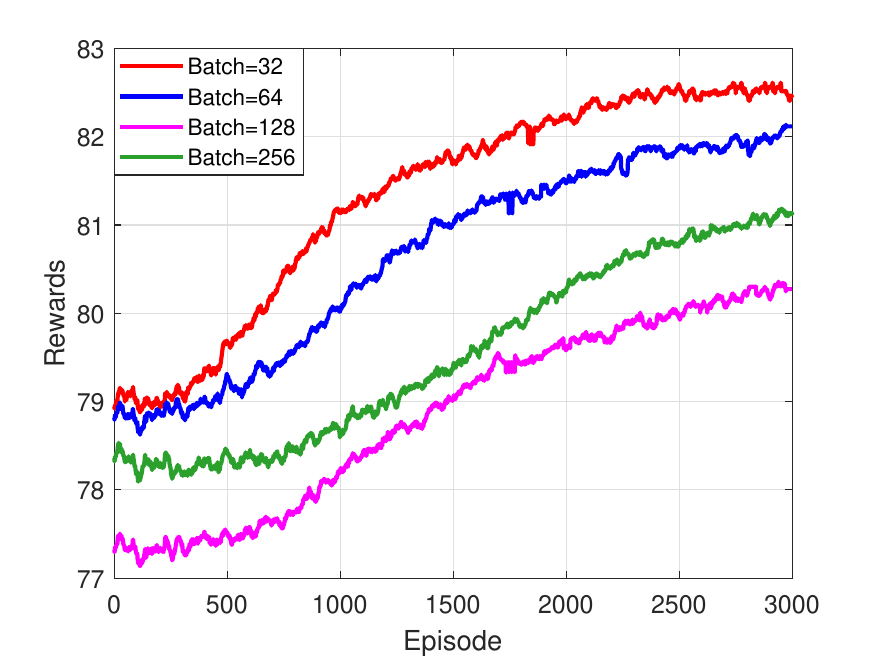}}
\caption{The comparison of convergence performance of different batch size.}
\label{fig:labe6}
\vspace{-0.65em}
\end{figure}

We conduct a detailed performance analysis of our proposed algorithm by examining the effects of various hyperparameters, including the discount factor $\gamma$, the clipping coefficient $\epsilon$, and the batch size for PPO training. 
First, we explore the impact of different $\gamma$ values (0.8, 0.9, 0.95, and 0.99) on performance.  As shown in Fig. \ref{fig:labe4}, the algorithm achieves the highest reward and demonstrates good convergence with $\gamma = $ 0.99, indicating that a higher discount factor is better for longer-term tasks by balancing immediate and future rewards. Next, Fig. \ref{fig:labe5} illustrates the effect of the clipping coefficient $\epsilon$ on performance. Testing values from 0.1 to 0.5 reveals that $\epsilon = 0.4$ yields the highest rewards, optimizing the balance between training stability and exploration capability. Finally, Fig. \ref{fig:labe6} examines different batch sizes. Our experiments show that a batch size of 32 consistently produces the highest rewards, striking an optimal balance between training stability and computational efficiency. Therefore, we select 
$\gamma = 0.99$, $\epsilon = 0.4$, and a batch size of 32 to enhance the PPO algorithm's performance and adaptability in dynamic environments.

\begin{figure}[!htbp]
\vspace{-0.5em}
\centerline{\includegraphics[width=0.40\textwidth]{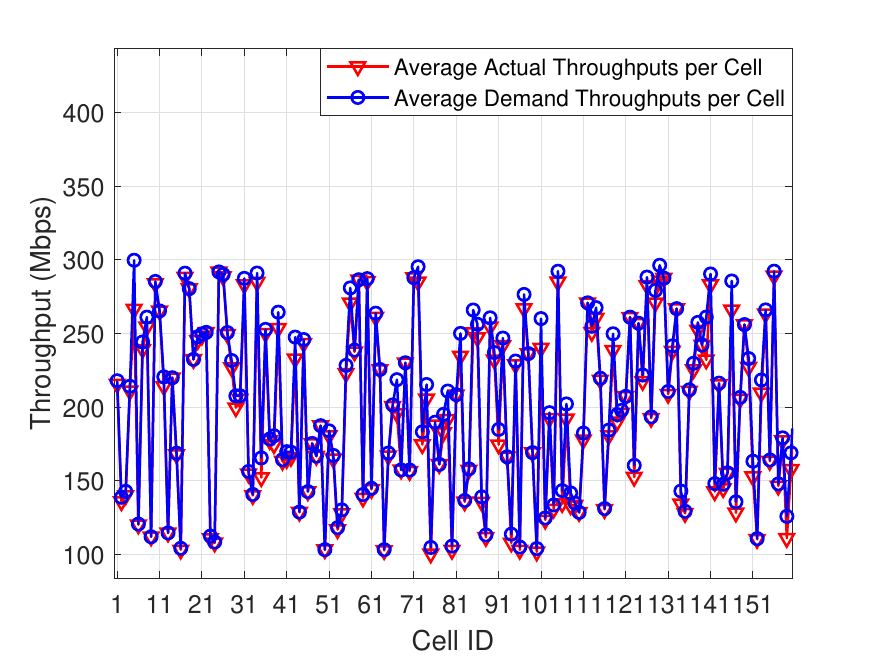}}
\caption{The performance of throughput and demand for each cell.}
\label{fig:7}
\vspace{-0.65em}
\end{figure}

Fig. \ref{fig:7} shows the demand and throughput comparison for 161 cells, where the cell traffic arrival rates are distributed from 100 Mbps to 300 Mbps. This illustration reveals significant differences in the ability of different cells to handle data traffic. Specifically, the maximum throughput of different cells varies by as much as 200 Mbps, indicating a significant imbalance in traffic demand between cells. However, the traffic requirements of most cells are effectively met, indicating that the proposed algorithm performs well in scheduling and resource allocation and can balance the load of each cell to a large extent.

\begin{figure}[!htbp]
\vspace{-0.5em}
\centerline{\includegraphics[width=0.40\textwidth]{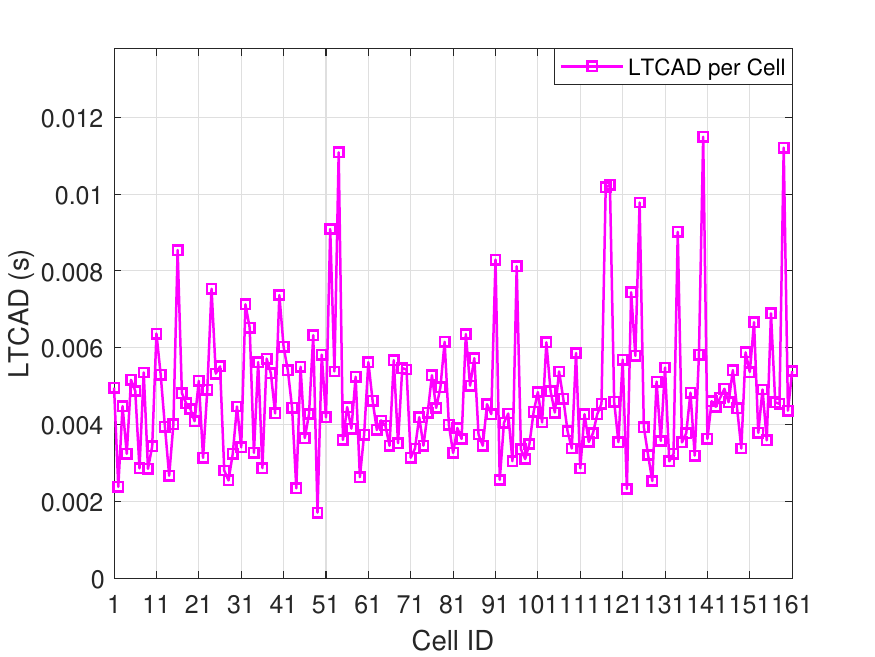}}
\caption{The performance of LTCAD for each cell.}
\label{fig:8}
\vspace{-0.65em}
\end{figure}

Fig. \ref{fig:8} shows the LTCAD of these 161 cells. Under the condition that the traffic arrival rate is still between 100 and 300 Mbps, although the latency of individual cells is higher, overall, the maximum latency of most cells does not exceed 6 milliseconds, and the LTCAD remains at about 4.6 milliseconds. This shows that the algorithm has excellent performance in optimizing system performance and can ensure low-latency data transmission while meeting high throughput requirements, thus improving the efficiency and user experience of the overall communication system.

\begin{figure}[!htbp]
\vspace{-0.5em}
\centerline{\includegraphics[width=0.40\textwidth]{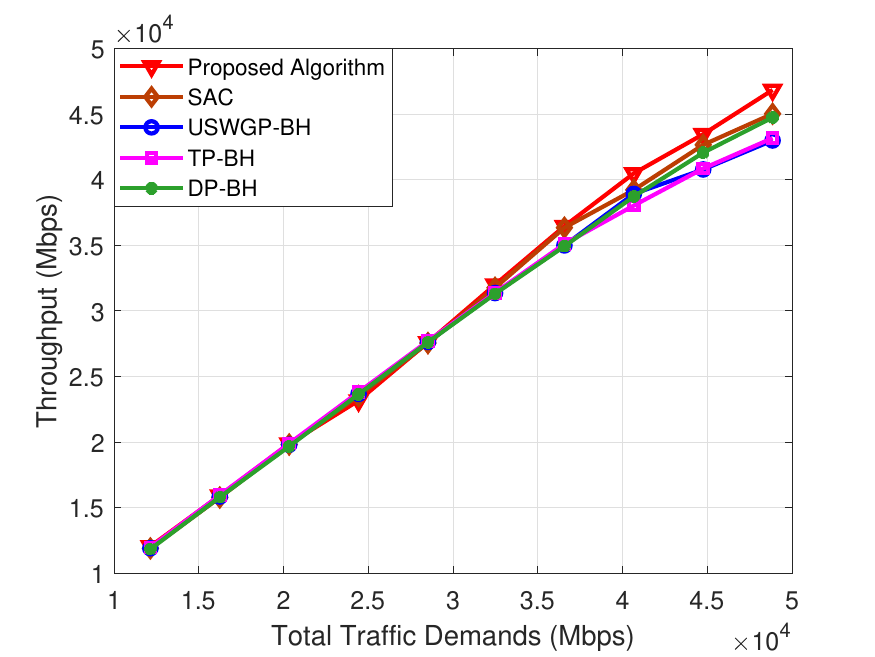}}
\caption{The long-term throughput versus total traffic demands of different methods.}
\label{fig:5}
\vspace{-0.65em}
\end{figure}

To further simulate the time-varying characteristics of data traffic, the traffic arrival rate for each cell is randomly distributed between 50 Mbps and 500 Mbps. The total traffic demand is defined as the sum of the traffic demands of the 161 cells.
Fig. \ref{fig:5} illustrates the relationship between long-term throughput and total traffic demand across ten different traffic arrival rate distributions for various methods. We evaluate ten distinct traffic arrival rate modes.
As shown in Fig. \ref{fig:5}, the performance of the proposed method is comparable to that of other algorithms under the first five traffic distributions. However, as the total traffic demand increases in the last five distributions, the proposed algorithm achieves higher throughput. Additionally, the throughput gap between the proposed algorithm and other algorithms widens with increasing traffic demand. Specifically, when the total traffic demand is 46834.6 Mbps, the proposed algorithm outperforms the SAC algorithm by 4.0\%, the USWGP-BH by 8.9\%, the TP-BH by 8.5\%, and the DP-BH by 4.6\%. 
Since the capacity of each beam is much greater than the traffic demand of each cell, all methods can meet the demand when the total traffic demand is low. However, as the total traffic demand increases, the proposed method outperforms the others by flexibly allocating limited beam resources to different cells.

\begin{figure}[!htbp]
\vspace{-0.5em}
\centerline{\includegraphics[width=0.40\textwidth]{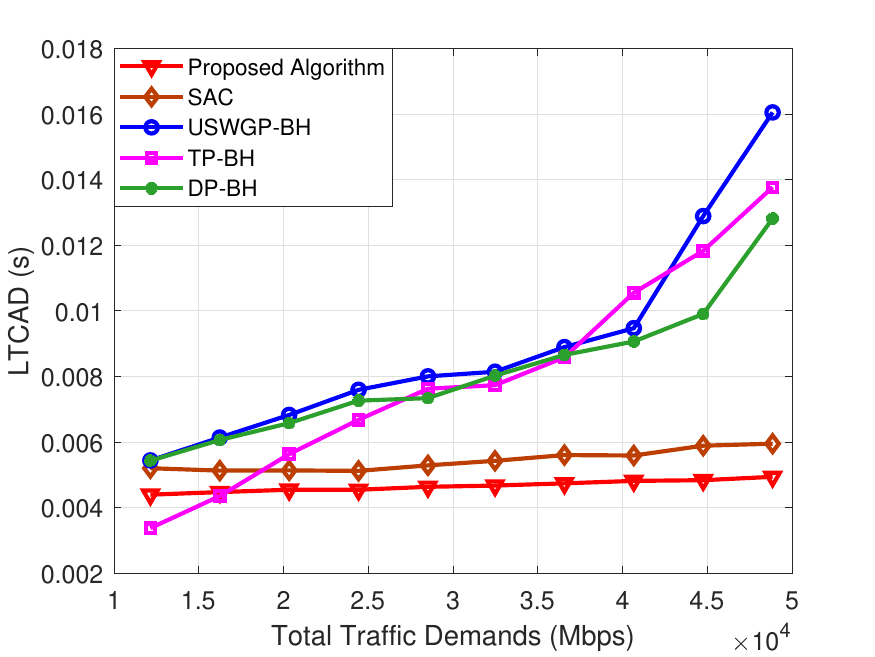}}
\caption{The LTCAD versus total traffic demands of different methods.}
\label{fig:6}
\vspace{-0.65em}
\end{figure}

To verify the performance in terms of LTCAD, we test the LTCAD performance in the same scenario, as shown in Fig. \ref{fig:6}.
As total traffic demand increases, the LTCAD for all methods also increases due to the limited beam resources' inability to meet the excessive traffic demand, leading to queue congestion.
When the total traffic demand grows, the mismatch between limited beam resources and uneven traffic demand makes it more challenging to guarantee low delay. In this context, the DP-BH algorithm outperforms TP-BH and USWGP-BH, mainly because DP-BH prioritizes serving the cells with the highest delay in each time slot, thereby more effectively reducing the overall delay.
When the total traffic demand is small, each beam can meet the needs of the destination cell, and the DP-BH algorithm prioritizes serving the cell with a long queue length, so its LTCAD is better than other algorithms.
The delay of the SAC algorithm remains relatively stable.
The proposed algorithm outperforms other algorithms, and this advantage becomes more pronounced as total traffic demand increases. Specifically, under maximum load conditions, the proposed algorithm reduces LTCAD by 20.5 \% compared to SAC, 69.2 \% to USWGP-BH, 64.1 \% to DP-BH, and 61.4 \% to TP-BH. This significant performance improvement demonstrates that the proposed algorithm has a clear advantage in resource allocation and delay management, enabling it to handle high loads and uneven traffic demands better. 

\section{Conclusion}
This paper investigates the beam scheduling and resource allocation problems in multi-NGSO BH scenarios. A DRL-based multi-NGSO BH algorithm is proposed to maximize network throughput and minimize the LTCAD. The algorithm fully utilizes the three degrees of freedom of beams in time, space, and power. It adopts joint optimization of beam illumination patterns and power allocation to achieve traffic-driven BH scheduling and resource allocation.
We have conducted multiple simulation experiments under different traffic demand scenarios and compared the performance of four benchmark BH algorithms. As the total traffic demand increases, the performance advantage of our proposed algorithm becomes more apparent. Under the heaviest traffic scenario, our algorithm improves the system throughput by up to 8.9 \% and reduces the LTCAD by 69.2 \%. In addition, the traffic demand of each cell is efficiently met. These results show that the proposed algorithm significantly reduces the LTCAD in time-varying traffic scenarios while improving system throughput, thereby optimizing overall system performance.


\small
\bibliographystyle{ieeetr}
\bibliography{reference}


\end{document}